\documentclass[final]{ws-mpla}

\usepackage[super,compress]{cite}
\usepackage{amsmath,amssymb,amsfonts}
\usepackage{bm,bbm}
\usepackage{graphicx}
\usepackage[english]{babel}
\usepackage[utf8]{inputenc}
\usepackage[OT2,T1]{fontenc}
\usepackage{silence} 
\usepackage{color}
\usepackage[usenames,dvipsnames]{xcolor}
\definecolor{ao(english)}{rgb}{0.0, 0.5, 0.0}
\usepackage[breaklinks,colorlinks=true,linkcolor=blue,citecolor=ao(english)]{hyperref}
\usepackage[hyphenbreaks]{breakurl}
\usepackage[usenames,dvipsnames]{xcolor}
\usepackage{pifont} 
\usepackage{makecell}
\usepackage{multirow}
\usepackage{arydshln} 

\let\normalcolor\relax  

\def\cob{\color{blue}}

\newcommand{\be}{\begin{equation}}
\newcommand{\ee}{\end{equation}}
\newcommand{\ba}{\begin{eqnarray}}
\newcommand{\ea}{\end{eqnarray}}
\def\bs{\begin{subequations}}
\def\es{\end{subequations}}
\def\a{\alpha}
\def\b{\beta}
\def\de{\delta}
\def\g{\gamma}
\def\la{\lambda}
\def\k{\kappa}

\def\om{\omega}

\def\s{\sigma}

\def\vp{\varphi}

\def\cD{\mathcal{D}}

\def\cG{\mathcal{G}}

\def\cK{\mathcal{K}}
\def\cL{\mathcal{L}}

\def\cP{\mathcal{P}}

\def\cR{\mathcal{R}}

\def\bE{\mathbbm{e}}

\def\ds{d_{\rm S}}
\def\dh{d_{\rm H}}
\def\dw{d_{\rm W}}

\def\p{\partial}

\def\B{\Box}

\def\cob{\color{blue}}

\newcommand{\Eq}[1]{(\ref{#1})}

\newcommand{\au}[2]{#1.~#2}
\newcommand{\book}[5]{\emph{#1} (#2, #5)}
\newcommand{\books}[4]{\emph{#1} (#2, #4)}
\newcommand{\procsinm}[5]{in \emph{#1}, eds.\ #2 (#3, #5)}
\newcommand{\oarX}[1]{\href{http://arxiv.org/abs/#1}{{\ttfamily\cob arXiv:#1}}}
\newcommand{\arX}[1]{\href{http://arxiv.org/abs/#1}{{\ttfamily\cob arXiv:#1}}}
\newcommand{\doin}[6]{\href{http://dx.doi.org/#1}{{\cob {\it #2 #3} {\bf #4}, #5 (#6)}}}
\newcommand{\doinn}[5]{\href{http://dx.doi.org/#1}{{\cob {\it #2} {\bf #3}, #4 (#5)}}}
\newcommand{\doij}[5]{\href{http://dx.doi.org/#1}{{\cob {\it #2} {\bf #3}, #4 (#5)}}}

\newcommand{\ndoinn}[5]{\href{#1}{{\cob {\it #2} {\bf #3}, #4 (#5)}}}
\newcommand{\tia}[1]{}

\def\lpl{\ell_{\rm Pl}}
\def\tpl{t_{\rm Pl}}
\def\rme{e}
\def\rmd{d}
\def\rmi{i}
\renewcommand{\leq}{\leqslant}
\renewcommand{\geq}{\geqslant}

\begin{document}


\title{Multifractional theories: an updated review}
\author{Gianluca Calcagni}
\address{Instituto de Estructura de la Materia, IEM-CSIC,\\ Serrano 121, 28006 Madrid, Spain\\ g.calcagni@csic.es}

\maketitle

\begin{abstract}
The status of multifractional theories is reviewed using comparative tables. Theoretical foundations, classical matter and gravity dynamics, cosmology and experimental constraints are summarized and the application of the multifractional paradigm to quantum gravity is discussed. We also clarify the issue of unitarity in theories with integer-order derivatives.
\end{abstract}

\date{March 4, 2021}




\



\section{Introduction}

In the attempt to unify the forces of Nature, several proposals of a quantum theory of gravitation have been flourished from the last quarter of the XX century until today \cite{Ori09,Fousp}. There is no unique answer to the question about how to quantize gravity consistently and, until empirical evidence of phenomena beyond general relativity is found, it is not possible to decide which theory, if any among the extant ones, describes more faithfully the physics at the frontier between gravity and quantum interactions.

Why are there different solutions to the problem of quantum gravity? It has to do with the way we picture ourselves the problem, as exemplified by the following quotation:
\begin{quote}
Let us take a real problem: The machines designed to pick tomatoes are damaging the tomatoes. What should we do? If we represent the problem as a faulty machine design, then the goal is to improve the machine. But if we represent the problem as a faulty design of the tomatoes, then the goal is to develop a tougher tomato \cite{Woo19}.
\end{quote}
In our context, the machine is perturbative quantum field theory (QFT), the tomato is classical gravity and the final product, canned tomatoes, is quantum gravity. Some theories of quantum gravity represent the issue of the non-renormalizability of perturbative quantum gravity as perturbative QFT being faulty and they recur to different machines applied to classical general relativity: the functional renormalization-group approach in asymptotic safety \cite{NiR,CPR,RSnax}, the canonical quantization in Ashtekar--Barbero variables in loop quantum gravity \cite{rov07,thi01}, or a path integral over triangulated geometries in causal dynamical triangulations \cite{AGJL4,Loll:2019rdj}. Other theories opt for keeping perturbative QFT as their machine while changing the tomato for an altogether different fruit, such as strings on a worldsheet \cite{BBSb,Zwi09} or fields on a group manifold as in group field theory \cite{Ori09,Fousp,GiSi}. Still other proposals content themselves with modifying the tomato just enough to make the perturbative-QFT machine working well enough not to crush the fruit: this is the case of the perturbative quantization of modified gravitational actions as in nonlocal quantum gravity \cite{Modesto:2017sdr} and multifractional theories, the topic of this paper.

Multifractional theories are classical and/or quantum field theories of gravity and matter characterized by a spacetime with \emph{multiscale} properties, i.e., the phenomena registered by clocks, rulers and detectors depend on the probed scale. This is a common feature of theories of quantum gravity \cite{tH93,Car09,fra1,Car17,MiTr} that usually emerges as a byproduct, while in multifractional spacetimes it is built in explicitly from the start. The way to do so is by modifying the integro-differential calculus defining the action, the dynamics, the line element, and so on. While it turns out to be difficult to improve the renormalizability of the gravitational interaction, this shift of paradigm from ordinary to \emph{anomalous} geometry opened a Pandora's box of conceptual insights and phenomenology that, on one hand, have tied together several loose strands that can contribute to a unified picture of quantum-gravity models and their capabilities and, on the other hand, has hopefully led modified gravity and quantum gravity closer to experiments.

The purpose of this paper is to offer an updated review on the topic of multifractional theories. The most complete review to date is Ref.\ \citen{revmu}, but while the latter discusses comprehensively the conceptual framework of these models and its development until early 2017, here we will concentrate on the general classification of the theories and their characteristics, stressing some of the advances made since then:
\begin{itemize}
\item The understanding of logarithmic oscillations as the manifestation of complex dimensions \cite{Calcagni:2017via}, with new applications to inflation \cite{Calcagni:2017via} and late-time cosmic acceleration \cite{Calcagni:2020ads}.
\item The construction of a stochastic or fuzzy version of multifractional spacetimes where the fractional power-law corrections to the measure and, hence, to lengths is reinterpreted as an intrinsic uncertainty on distance measurements \cite{Amelino-Camelia:2017pdr,Calcagni:2017jtf}.
\item The construction of black-hole solutions \cite{Calcagni:2017ymp}.
\item New observational constraints from the Standard Model of electroweak and strong interactions \cite{frc16} and new applications to the field of gravitational waves (GWs), in particular regarding the luminosity distance of standard sirens \cite{Calcagni:2019ngc} and the primordial stochastic GW background \cite{Calcagni:2020tvw}.
\item The construction of multifractional derivatives \cite{mf0} and of the long-sought theories with fractional differential operators \cite{mf1}.
\end{itemize}
Since we do not aim at covering all the theoretical aspects considered in Ref.\ \citen{revmu}, the present review may be seen as a complement to that publication. A forthcoming textbook will give a longer, in-depth introduction to multiscale and multifractional theories \cite{book}.

Section \ref{sec2} introduces the main features of the multifractional geometry of spacetime and the classification of multifractional theories as we understand them now. Classical gravity and cosmology in each theory is discussed in Sec.\ \ref{sec3}, while QFT and quantum gravity are reviewed in Sec.\ \ref{sec4}. We give a perspective on future research on the subject in Sec.\ \ref{sec5}.


\section{Spacetime geometry of multifractional theories}\label{sec2}

The geometry of spacetime defining multifractional theories can be read out from the prototype massless scalar-field action
\ba
&&S=\int\rmd^Dq(x)\,\left[\frac12\phi\cK\phi-V(\phi)\right],\\
&&\rmd^Dq(x)=\prod_\mu\rmd q^\mu(x^\mu)=\rmd^Dx\,\prod_\mu v_\mu(x^\mu)=:\rmd^Dx\,v(x)\,,
\ea
where $D$ is the number of topological dimensions ($D=4$ for physical models), $v(x)$ is a measure weight that depends on the spacetime coordinates $x^\mu$, $\mu=0,1,\dots,D-1$, $\cK$ is the kinetic operator depending on the derivatives ``$\mathbf{D}$''  defining the theory, and $V$ is a potential including nonlinear interactions. Each theory is named with a label $T_1$, $T_v$, $T_q$, \dots that roughly summarizes its differential structure. 

In this section we do not consider gravity, so that the spacetime metric is the Minkowski metric
\be
\eta_{\mu\nu}={\rm diag}(-,+,\cdots,+)_{\mu\nu}\,.
\ee
Therefore, here covariant operators such as the d'Alembertian $\B=\eta^{\mu\nu}\p_\mu\p_\nu$ are defined with zero affine connection. Lorentz indices are contracted with the Minkowski metric in Einstein convention, while they are not contracted in one-directional expressions such as $q^\mu(x^\mu)$.

The integro-differential structure and geometry of multifractional theories is compared in Tables \ref{tab1} and \ref{tab2} for theories with integer-order operators and in Table \ref{tab3} for theories with fractional operators.

\begin{table}
\tbl{Characteristics of spacetime geometry of multifractional theories with integer-order operators on Minkowski spacetime.  Acronyms: ultraviolet (UV), infrared (IR), discrete scale invariance (DSI). Ultra-IR means scales much beyond particle-physics scales, for instance cosmological.}
{\makebox[\textwidth][c]{
\begin{tabular}{|c||c|c|c|}\hline
  & $\bm{T_1}$ & $\bm{T_v}$ & $\bm{T_q}$ \\
{\bf Theory} & \makecell{{\bf ordinary derivatives}} & \makecell{{\bf weighted derivatives}} & $\bm{q}${\bf-derivatives}\\\hline\hline
\makecell{General\\ measure\\ $\displaystyle\prod_\mu \rmd q^\mu(x^\mu)$}  & \multicolumn{3}{c|}{\makecell{$\begin{aligned}
q^\mu(x^\mu)&=\sum_{l}\frac{\ell_l}{\a_{l}}{\rm sgn}(x^\mu)\left|\frac{x^\mu}{\ell_l}\right|^{\a_{l}} F_l(x^\mu)\\ F_l(x^\mu)= A_{l,0}+\sum_{n=0}^\infty A_{l,n}&\cos\left(n\om_{l}\ln\left|\frac{x^\mu}{\ell_l}\right|\right)+B_{l,n}\sin\left(n\om_{l}\ln\left|\frac{x^\mu}{\ell_l}\right|\right)
 &\end{aligned}$\\ 
Most general measure with factorizable coordinate dependence \cite{frc2,first,revmu,Calcagni:2017via}}}\\\hline
\makecell{Isotropic\\ measure\\ (two scales)} & \multicolumn{3}{c|}{\makecell{$q^\mu(x^\mu)\simeq \left|\frac{x^\mu}{\ell_*}\right|^{\a_*-1}F_*(x^\mu)+x^\mu+\left|\frac{x^\mu}{\ell_{\rm c}}\right|^{\a_{\rm c}-1}F_{\rm c}(x^\mu)$ \\ 
Minimal measure allowing for UV/subatomic ($\ell\lesssim\ell_*$)\\ and ultra-IR/cosmological ($\ell\gtrsim\ell_{\rm c}$) modifications of geometry \cite{Calcagni:2020ads}}}\\\hline
\makecell{Derivatives\\ $\mathbf{D}$} & $\p_\mu:=\dfrac{\p}{\p x^\mu}$ & $\begin{aligned}&\qquad\quad{}^\b\cD_\mu:=v^{-\b}\p_\mu(v^\b\,\cdot\,)\\ &\b=\begin{cases}\frac12\,\,\textrm{scalar \cite{frc3,frc6}, vector \cite{frc8,frc12,frc13}}\\ \frac{2}{D-2}\,\,\textrm{rank-2 tensor \cite{frc6,frc11}}\end{cases}\end{aligned}$\!\!\!\!\!\!\!\! & $\begin{aligned}\p_{q^\mu}&:=\dfrac{\p}{\p q^\mu(x^\mu)}\\ &=\dfrac{1}{v_\mu(x^\mu)}\dfrac{\p}{\p x^\mu}\end{aligned}$\\\hline
\makecell{$\cK$\\ (one scale)} & \makecell{$\vphantom{\displaystyle\int}\eta^{\mu\nu}\p_\mu\p_\nu$\, or \cite{frc2}\\ $\eta^{\mu\nu}\!\left(\p_\mu\p_\nu+\frac{\p_\mu v}{v}\p_\nu\right)$} & $\eta^{\mu\nu}{}^\b\cD_\mu{}^\b\cD_\nu$ & $\eta^{\mu\nu}\p_{q^\mu}\p_{q^\nu}$ \\\hline
\makecell{$\dh$\\ (isotropic measure)\\ (two scales)} & \multicolumn{3}{c|}{\!\!$\begin{aligned}\!\!\textrm{ultra-IR:}&\quad D\a_{\rm c}>D\\
\textrm{IR:}&\quad D\\ \textrm{UV:}&\quad D\a_*<D\end{aligned}$\!\!} \\\hline
\makecell{$\dh^k$\\ (isotropic measure)\\ (two scales)} & \multicolumn{3}{c|}{$\simeq\dh$} \\\hline
\makecell{$\ds$\\ (isotropic measure)\\ (two scales)} & \multicolumn{2}{c|}{$D$} & $\begin{aligned}\textrm{ultra-IR:}&\quad D\a_{\rm c}>D\\
\textrm{IR:}&\quad D\\ \textrm{UV:}&\quad D\a_*<D\end{aligned}$ \\\hline
Multifractal & \multicolumn{2}{c|}{No \cite{frc7}} & Yes \cite{frc7} \\\hline
\makecell{Measure\\ symmetries} & \multicolumn{3}{c|}{\makecell{$\bullet$ IR: Ordinary Poincar\'e symmetry \\ $\bullet$ At plateaux $\a\simeq{\rm const}$: DSI\quad $F_l(\la_l^n x)=F_l(x)$, $\la_l=\exp(-2\pi/\om_l)$}} \\\hline
\makecell{Lagrangian\\ symmetries} & \makecell{Ordinary Poincar\'e\\ symmetry \cite{fra2}} & \makecell{Weighted Poincar\'e\\ symmetry \cite{frc6,frc13,revmu}\\ $\bullet$ Free field theory:\\ ordinary algebra \\ $\bullet$ Interacting theory:\\ deformed algebra} & \makecell{$q$-Poincar\'e\\ symmetry \cite{frc1,frc2,Calcagni:2016ivi,revmu}} \\\hline
Integer frame & No \cite{revmu} & Yes, by field redefinition \cite{frc6,revmu} & \!\! Yes, in $q$-coordinates \cite{revmu}\!\! \\\hline
\end{tabular}\label{tab1}}}
\end{table}

\begin{table}
\tbl{Upper bound $\ell_*<\ell_*^{\rm max}$ for the UV scale of multifractional theories with integer-order operators. Cells report $\ell_*^{\rm max}$ in meters. For reference, $\lpl\approx 10^{-35}\,{\rm m}$. Empty cells correspond to cases not studied yet, while cells with a \ding{55} sign indicate that the constraint is not applicable because trivial or too weak. Bounds with a ${}^\dagger$ sign can be avoided (see Sec.\ \ref{sec:exper}). Acronyms: quantum electrodynamics (QED), gravitational waves (GWs), gamma-ray bursts (GRBs), cosmic microwave background (CMB).}{
\begin{tabular}{|c||c|c|c|}\hline
  & $\bm{T_1}$ & $\bm{T_v}$ & $\bm{T_q}$ \\
{\bf Theory} & \makecell{{\bf ordinary}\\ {\bf derivatives}} & \makecell{{\bf weighted}\\ {\bf derivatives}} & $\bm{q}${\bf-derivatives}\\\hline\hline
Muon lifetime \cite{frc16} & & $10^{-21}$ & $10^{-5}$ \\
Tau lifetime \cite{frc16} & & $10^{-19}$ & $10^{-10}$ \\
$K^{0}-\bar{K}^{0}$ transitions \cite{frc16} & & $10^{-19}$ & $10^{-6}$ \\
Lamb shift \cite{frc12,frc13} & & $10^{-14}$ & $10^{-13\,\dagger}$ \\
$\a_{\rm QED}$ measurements \cite{frc13} & & $10^{-18}$ & \ding{55} \\
$\frac{\Delta\a_{\rm QED}}{\a_{\rm QED}}$ quasars \cite{frc8} & & $10^{+20}$ & \ding{55} \\
\makecell{GW dispersion relation \cite{Yunes:2016jcc,qGW,revmu}} & \ding{55} & \ding{55} & $10^{-14\,\dagger}$ \\
\makecell{GW luminosity distance \cite{Calcagni:2019ngc}} & \ding{55} & \ding{55} & \ding{55} \\
GRBs \cite{qGW,revmu} & & \ding{55} & $10^{-30\,\dagger}$ \\
Cherenkov radiation \cite{revmu} & & \ding{55} &  $10^{-49\,\dagger}$ \\
CMB primordial spectra \cite{frc14} & & & \ding{55} \\\hline
\end{tabular}\label{tab2}}
\end{table}

\begin{table}
\tbl{Characteristics of spacetime geometry of multifractional theories with fractional operators on Minkowski spacetime. Empty cells correspond to topics not studied yet.}
{\makebox[\textwidth][c]{
\begin{tabular}{|c||c|c|c|c|}\hline
  & $\bm{T[\p+\p^\g]}$ & $\bm{T[\p^{\g(\ell)}]}$ & $\bm{T[\B+\B^\g]}$ & $\bm{T[\B^{\g(\ell)}]}$ \\
{\bf Theory} & \multicolumn{2}{c|}{{\bf fractional derivatives}} & \multicolumn{2}{c|}{{\bf fractional d'Alembertian}}\\\hline\hline
\makecell{General\\ measure\\ $\displaystyle\prod_\mu \rmd q^\mu(x^\mu)$}  & \multicolumn{4}{c|}{Ordinary measure\, $q^\mu(x^\mu)=x^\mu$} \\\hline
\makecell{Derivatives\\ $\mathbf{D}$\\ (one scale)} & \makecell{$\vphantom{\displaystyle\int}\cD_\mu:=\ell_*^{1-\g}\p+\cD^{\g}$\\
$\vphantom{\displaystyle\int}\cD^\g_\mu:=\frac12\left({}_\infty\p^\g_\mu-{}_\infty\bar \p^\g_\mu\right)$\\ ${}_\infty\p^\g$: Liouville derivative\\ ${}_\infty\bar\p^\g$: Weyl derivative \cite{mf0,mf1}} & $\cD^{\g(\ell)}_\mu$ & \multicolumn{2}{c|}{$\p_\mu$}  \\\hline
\makecell{$\cK$\\ (one scale)} & \makecell{$\eta^{\mu\nu}\cD_\mu\cD_\nu$\\ $\vphantom{\displaystyle\int}\g>1$} & $\eta^{\mu\nu}\cD^{\g(\ell)}_\mu\cD^{\g(\ell)}_\nu$ & \makecell{$\ell_*^{2-2\g}\B+\B^\g$\\ $\vphantom{\displaystyle\int}\g>1$\\ $\B^\g$: fractional\\ d'Alembertian \cite{mf1}} & $\B^{\g(\ell)}$ \\\hline
$\dh$ & \multicolumn{4}{c|}{$D$} \\\hline
\makecell{$\ds$\\ (two scales)} & $\begin{aligned}\textrm{ultra-IR:}&\quad {D}/{\g_{\rm c}}\\
\textrm{IR:}&\quad D\\ \textrm{UV:}&\quad {D}/{\g_*}\end{aligned}$ & $\dfrac{D}{\g(\ell)}$ & $\begin{aligned}\textrm{ultra-IR:}&\quad {D}/{\g_{\rm c}}\\
\textrm{IR:}&\quad D\\ \textrm{UV:}&\quad {D}/{\g_*}\end{aligned}$ & $\dfrac{D}{\g(\ell)}$ \\\hline
Multifractal & & & & \\\hline
\makecell{Lagrangian\\ symmetries} & \makecell{None exact \cite{revmu,mf1}\\ $\bullet$ IR: ordinary Poincar\'e\\ symmetry\\ $\bullet$ At plateaux $\g\simeq{\rm const}$:\\ fractional Poincar\'e\\ symmetry \cite{frc2,revmu}} & \makecell{Fractional Poincar\'e\\ symmetry of\\ variable order \cite{frc2,mf1}} & \multicolumn{2}{c|}{\makecell{Ordinary Poincar\'e\\ symmetry \cite{mf1}}} \\\hline
Integer frame & \multicolumn{4}{c|}{No \cite{mf1}} \\\hline
\makecell{Constraints\\ on $\ell_*$} & & & & \\\hline
\end{tabular}\label{tab3}}}
\end{table}

\pagebreak


\subsection{About spacetime measure}

The time and length scales $\ell_l\geq\lpl$ are by assumption fundamental scales of spacetime geometry. This is why these spacetimes are called multiscale \cite{Calcagni:2016edi}. For only one fundamental UV scale $\ell_*$, quantum-gravity arguments suggest to identify it with the Planck scale $\ell_*=\lpl$, together with setting $\a_*=1/2$ or $\a_*=1/3$ \cite{Calcagni:2017jtf}. In the following, we will not set these parameters to any fixed value.

In the first papers on multiscale spacetimes, the spacetime measure $\rmd^Dq(x)$ appearing in Table \ref{tab1} had been proposed as a profile naturally possessing critical exponents \cite{fra1,fra2,frc2} and the discrete symmetry typical of deterministic fractals \cite{fra4,frc1,frc2}. The idea was that, on one hand, multiscale phenomena are universally described by critical exponents and that, if the geometry of spacetime had one or more fundamental scales, then its measure should also be of the form of a generalized polynomial with different exponents $\a_l$. This first argument led to the power-law dependence $\sim |x|^\a$ of $q(x)$. On the other hand, deterministic multifractals (i.e., fractals which are exactly self-similar) are a special case of multiscale systems where the critical exponents $\a\pm\rmi\om$ are complex and the complex part is associated with a discrete scale invariance (DSI) \cite{Sor98,NLM} and a length scale $\ell_\infty$ appearing inside the logarithms to make their argument $x/\ell_\infty$ dimensionless. This led to the log-oscillatory dependence in $q(x)$. However, in Refs.\ \citen{first,revmu} a much stronger result was proven, namely, that this spacetime measure not only obeys the above two universal features of exactly self-similar multiscale systems, but it is also the most general measure under three assumptions: (i) spacetime is a continuum, (ii) the ordinary Lebesgue measure must be recovered in some regime which is reached ``slowly enough'' (i.e., via a flat asymptote), and (iii) the measure is factorizable in the coordinates. This result goes under the name of \emph{flow-equation theorem}.\footnote{Assumption (iii) can be relaxed and one can apply the theorem directly to the Hausdorff and the spectral dimension \cite{first,revmu}.} According to the same theorem, the derivation of the log-oscillatory part of the measure from the pairing of complex conjugate power laws leads to identify the two scales $\ell_\infty=\ell_*$, thus reducing the number of free parameters in the measure \cite{Calcagni:2017jtf}.

Multifractional spacetimes are by definition multiscale spacetimes with a factorizable measure \cite{Calcagni:2016edi}. The assumption of factorizability is made to drastically simplify calculations. This of course breaks Poincar\'e symmetries. Theories with more symmetries such as spatial rotations have been considered \cite{fra1,fra2,fra3}, but they are much more difficult to handle and it is preferable to give up Poincar\'e symmetries altogether and recover them at large scales \cite{frc2,frc6,Calcagni:2012kf,Calcagni:2016ivi}.

All the parameters $\a_l=\a_{l,\mu}$, $\ell_l=\ell_{l,\mu}$, $A_{l,n},B_{l,n}=A_{l,n,\mu},B_{l,n,\mu}$, $\om_l=\om_{l,\mu}$ in the general measure can be different for different directions, but in Table \ref{tab1} we wrote down a simplified ``isotropic'' version where the index $\mu$ has been omitted everywhere. However, it is not uncommon to consider geometries where only the time direction or only the spatial directions are multifractional. 

The $n$-dependence of the amplitudes $A_{l,n}$ and $B_{l,n}$ has been worked out in Ref.~\citen{Calcagni:2017via} looking at the typical dependence found in critical, complex, and fractal systems, and it is such that the amplitudes decrease exponentially or as a power law in the order $n$ of the harmonic:
\be\label{Aa}
A_n=a_n \frac{\rme^{-c n}}{n^u},\qquad B_n=b_n \frac{\rme^{-c n}}{n^u},
\ee
where $a_n$, $b_n$, $c$ and $u$ are constant in the deterministic version or view of the measure, while $a_n$ and $b_n$ are random variables in the so-called stochastic view \cite{Amelino-Camelia:2017pdr,Calcagni:2017jtf}, where the fractional corrections to the ordinary measure are stochastic fluctuations around the zero mode that make spacetime fuzzy. 


\subsection{About theories with fractional operators (I)}

The expressions for $\cK$ with explicit scale-dependence in Table \ref{tab3} (theories $T[\p+\p^\g]$ and $T[\B+\B^\g]$) are examples in the presence of only one scale $\ell_*$, but they can be generalized immediately to two or more scales.

\begin{figure}
\centering
\includegraphics[width=11cm]{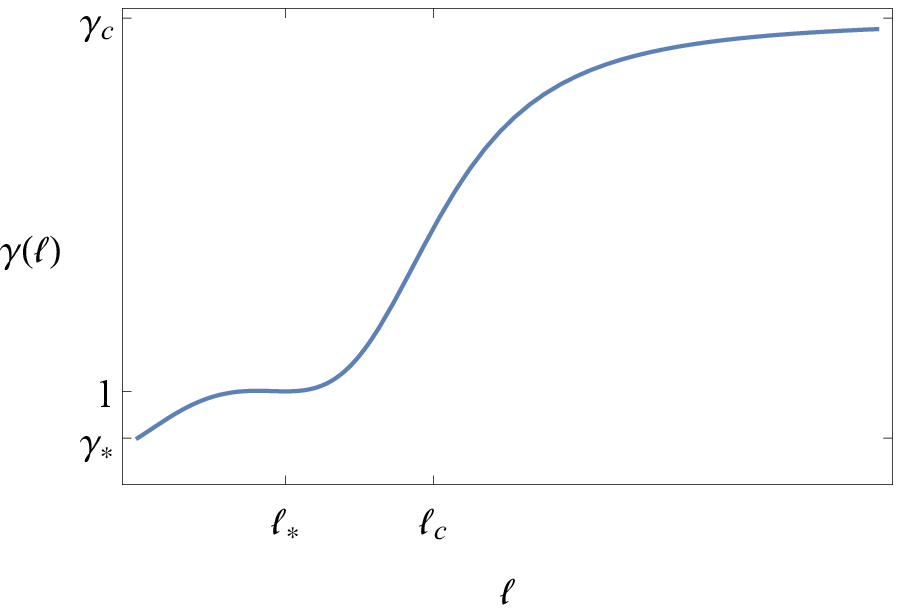}
\caption{\label{fig1} Example \Eq{Kellg3} of two-scale profile for the variable fractional order $\g(\ell)$ in the theories $T[\p^{\g(\ell)}]$ and $T[\B^{\g(\ell)}]$.}
\end{figure}

The profile $\g(\ell)$ in the theories $T[\p^{\g(\ell)}]$ and $T[\B^{\g(\ell)}]$ can be chosen to reproduce the asymptotic behavior of the spectral dimension $\ds$ shown in Table \ref{tab3}. One such profile could be (Fig.\ \ref{fig1})
\be\label{Kellg3}
\g(\ell)=\frac{\g_*-1+[(\ell_*/(\ell-\ell_*)]^2+(\ell_{\rm c}/\ell)^2\g_{\rm c}}{[(\ell_*/(\ell-\ell_*)]^2+(\ell_{\rm c}/\ell)^2}\,,
\ee
where $\ell_*\ll \ell_{\rm c}$. Other single and two-scale profiles can be found in Refs.\ \citen{frc2,fra6,frc4,Calcagni:2020ads,mf1}. The fact that these profiles are chosen \emph{ad hoc} can be considered as a weakness of these theories because it introduces an element of arbitrariness that, to date, we are unable to constrain with theoretical arguments. However, the payback is noteworthy because it may allow us achieve unitarity and renormalizability at the same time, something problematic for the theories $T[\p+\p^\g]$ and $T[\B+\B^\g]$ \cite{mf1}.

In all the theories with fractional operators in Table \ref{tab4}, the ordinary Lebesgue measure has been chosen because the generalized polynomial of the other theories is not necessary to improve renormalizability, but this assumption can be relaxed. In the cases with fractional derivatives $T[\p+\p^\g]$ and $T[\p^{\g(\ell)}]$, call the generalized theories with multifractional measure
\be\label{tt1}
T_\a[\p+\p^\g]\quad {\rm and}\quad T_\a[\p^{\g(\ell)}]\,.
\ee
In particular, there are strong similarities between the theories
\be\label{tt2}
T_\a[\p+\p^\a]\quad {\rm and}\quad T_\a[\p^{\a(\ell)}]
\ee
and the theory with $q$-derivatives, since the scaling of the $q$-derivative is the same as the scaling of fractional derivatives. This correspondence, which has not been explored yet, is indicated as \cite{revmu}
\be\label{cong}
T_{\g=\a}\cong T_q\,,
\ee
and it could mean that the observational constraints found for $T_q$ could be applied also to, or be very similar to those for, the theories \Eq{tt2}.


\subsection{About dimensions}

The Hausdorff dimension $\dh$ of spacetime is defined as minus the scaling of the position-dependent part of the spacetime measure,
\be
\dh:=-[\rmd^Dq(x)]_x=-[\rmd^Dx\,v(x)]_x=D-[v(x)]_x\,,
\ee
where $[x^\mu]=-1$, while the Hausdorff dimension $\dh^k$ of momentum space is the scaling of the momentum-dependent part of the momentum-space measure,
\be
\dh^k:=[\rmd^Dp(k)]_k=[\rmd^Dk\,w(k)]_k=D+[w(k)]_k\,,
\ee
where $[k^\mu]=1$ and usually $w(k)\neq v(k)$. Notice the specification that the scaling is the one of the variable part of the measure. For instance, by definition $[v]=0$, but its position-dependent part in the UV has $[v]_x=[x_0^{\a-1}\cdots x_{D-1}^{\a-1}]=D(1-\a)$. The momentum-space measure for $T_1$, $T_v$ and $T_q$ is discussed, respectively, in Refs.\ \citen{frc3}, \citen{frc3} and \citen{frc11}. Here we complete the discussion by determining the Hausdorff dimension of momentum space. For the theories $T_1$, $T_v$ and $T_q$, only if $p(k)=1/q(1/k)$ can one define an invertible momentum transform where the basis $\bE(k,x)$ of eigenfunctions of the kinetic operator $\cK$ are symmetric in $x$ and $k$ at any given plateau in dimensional flow (i.e., intervals of scales where the dimension is approximately constant) \cite{frc11}. This implies that $\dh^k=\dh$ at any plateau in dimensional flow.

The Hausdorff dimension of spacetime also has an imaginary part $d_{{\rm H}\mathbbm{C}}$ that coincides with the frequency of the log-oscillations \cite{Calcagni:2017via}:
\be
\om=d_{{\rm H}\mathbbm{C}}\,.
\ee
Since the expression $q(x)$ is real-valued, there is no physical problem with having complex dimensions and, in fact, we can even observe them in principle, for instance, as a modulation in the CMB primordial spectrum \cite{Calcagni:2017via} or as a cosmic acceleration at late times \cite{Calcagni:2020ads}. The peculiarity of DSI is that it is a UV symmetry that affects the IR even when broken. This departure from the usual UV/IR dichotomy happens because DSI is characterized by infinitely many scales $\la_l^{\pm n}\ell_*$ spanning all ranges.

The spectral dimension is related to the momentum space of the model. Consider the Schwinger representation of the Green function
\ba
\hspace{-.9cm}&&G(x,x') =-\int\rmd^Dp(k)\,\frac{\bE(k,x)\bE(k,x')}{\cK(k)}=\int_0^{+\infty}\rmd\s(\ell)\,P(x,x';\ell),\label{scw}\\
\hspace{-.9cm}&&P(x,x';\ell)=\int\rmd^Dp(k)\,\bE(k,x)\bE(k,x')\,\rme^{\s(\ell)\cK(k)}.
\ea
The spectral dimension is 
\be\label{ds}
\ds:=-\frac{\rmd\ln\cP(\ell)}{\rmd\ln\ell}\,,\qquad \cP(\ell):=\frac{\int\rmd^Dq(x)\,P(x,x;\ell)}{\int\rmd^Dq(x)}\,,
\ee
where $\cP(\ell)$ is called return probability. In multifractional theories, the basis functions $\bE$ are factorizable in the coordinates \cite{frc3,frc14} and we can write
\be\label{facte}
\bE(k,x)=\prod_\mu \bE_\mu(k_\mu, x_\mu)\,.
\ee
For the theory $T_v$ with weighted derivatives, the basis $\bE$ is \cite{frc3}
\be\label{tve}
T_v\,:\qquad \bE(k,x)=\prod_\mu \frac{\rme^{\rmi k_\mu x^\mu}}{\sqrt{2\pi w_\mu(k^\mu) v_\mu(x^\mu)}}=\frac{1}{(2\pi)^\frac{D}{2}}\frac{\rme^{\rmi k\cdot x}}{\sqrt{w(k)\,v(x)}}\,,
\ee
which are eigenfunctions of the operator $\cK=\B+(\p_\mu v/v)\p^\mu$ in the corresponding column in Table \ref{tab1}: $\cK\bE(k,x)=-k^2\bE(k,x)$. Since $\s(\ell)=\ell^2$ in this theory, after defining the dimensionless variable $y=\ell k$, one has $\cP(\ell)=[\int\rmd^Dx/\int\rmd^Dx\,v(x)]C\ell^{-D}$, where $C$ is a constant. Therefore, $\ds=D$ \cite{revmu}, which is the result obtained in Ref.~\citen{frc7} setting the parameters $\b$ and $\nu$ therein to their natural value $\b=1=\nu$. The theory $T_1$ follows the same dimensional flow \cite{frc7}. 

For the theory $T_q$ with $q$-derivatives, one has \cite{frc14}
\be\label{tqe}
T_q\,:\qquad \bE(k,x)=\prod_\mu\frac{\rme^{\rmi p_\mu(k^\mu) q^\mu(x^\mu)}}{\sqrt{2\pi}}=\frac{1}{(2\pi)^\frac{D}{2}}\,\rme^{\rmi p(k)\cdot q(x)}\,,
\ee
and $\cK\bE(k,x)=-p^2(k)\bE(k,x)$. Here $\s(\ell)=q^2(\ell)$ and, therefore, $\cP(\ell)$ scales as $\ell^{-D\a}$ in the UV or at any other plateau. The spectral dimension for the theories with fractional operators was calculated in Ref.\ \citen{mf1}, where some profiles for $\g(\ell)$ in the theories $T[\p^{\g(\ell)}]$ and $T[\B^{\g(\ell)}]$ were also proposed (see below).

All multifractional spacetimes are multiscale but not all are also multifractal. Multifractal spacetimes are such that the spectral and Hausdorff dimension are related to each other by a fixed relationship $\dw=2\dh/\ds$ that also involves the so-called walk dimension $\dw$, which is calculated independently \cite{Calcagni:2016edi}. Without entering into details, there is mathematical evidence that $\ds=\dh^k$ for fractals \cite{Akkermans:2010dz} and $T_q$ is the only theory among those with integer-order derivatives that satisfies this property.


\subsection{About symmetries}

Continuous symmetries are classified as ordinary or deformed depending on whether their generators are ordinary or not. However, these generators can satisfy the ordinary symmetry algebra in some cases, such as the free scalar field theory with weighted derivatives \cite{frc6}. The absence of action symmetries in the theory $T_1$ is responsible for the absence of a mathematical integer frame or picture \cite{revmu} where one can simplify the dynamics and make it superficially identical, or at least very similar, to standard mechanics or field theory. 

Concerning discrete symmetries in a QFT context, both $T_v$ and $T_q$ are CPT invariant (charge conjugation, parity and time reversal) \cite{frc13}.


\subsection{About experimental bounds}\label{sec:exper}

The simplified two-scale measure has been used to get experimental bounds on the scales $\ell_*$ and $\ell_{\rm c}$ from, respectively, particle-physics and cosmological observations, with or without log-oscillations (Table \ref{tab2}).

The upper bounds on the UV scale $\ell_*$ in Table \ref{tab2} are obtained for $\a_*\ll 1$ and are the weakest possible. They tighten progressively when $\a_*$ increases from 0 to 1 \cite{frc8,frc12,frc13,qGW,revmu,frc16}. For $T_v$ and $T_q$, there also exist constraints from the CMB black-body spectrum for a fixed $\a_*$, which are of the same order of magnitude as the bounds from particle physics \cite{frc14}.

In Table \ref{tab2}, bounds with a dagger (${}^\dagger$) are avoided in the stochastic view of the theory $T_q$ \cite{Amelino-Camelia:2017pdr,Calcagni:2017jtf}. If the zero mode in the measure vanishes, $A_0=0$, fractional corrections cancel out in average and the strongest bounds in Table \ref{tab2}, as well as the bounds from the CMB black-body spectrum \cite{frc14}, cease to be \cite{Calcagni:2017jtf}.

Bounds on $\a_*$, $\a_{\rm c}$ and $\ell_{\rm c}$, as well as constraints on the log-oscillation amplitudes, are also available:
\begin{itemize}
\item $\a_*<0.47$ in the stochastic view of the theory $T_q$, according to limits on the strain noise in present and future GW interferometers \cite{Calcagni:2019ngc}. In the deterministic view of the same theory and in the presence of one harmonic in the oscillatory part of the measure, $\a_*\lesssim 0.1\!-\!0.6$ if inflationary scales include those of the UV regime of the theory \cite{frc14}.
\item $A,B<0.4$ when $\a_*=1/2$ and the measure has only one harmonic with amplitudes $A$ and $B$, according to CMB constraints on inflation in the deterministic view of the theory $T_q$ \cite{frc14}.
\item $\a_{\rm c}\approx 3.8$ and $t_{\rm c}=\tpl\ell_{\rm c}/\lpl>3.9\,t_0$, where $t_0\simeq H_0^{-1}$ is the age of the universe, in the theory $T_v$ with many harmonics in the measure, from late-time measurements of the accelerated expansion of the universe \cite{Calcagni:2020ads}.
\end{itemize}


\section{Classical gravity in multifractional theories}\label{sec3}

Gravity in multifractional spacetimes is described by an action
\be
S=\frac{1}{2\k^2}\int\rmd^Dq(x)\,\sqrt{|g|}\,\cL_g+S_{\rm matter}\,,
\ee
where $\k^2=8\pi G$, $G$ is Newton's constant, $g$ is the determinant of the metric $g_{\mu\nu}$, $\cL_g$ is the gravitational Lagrangian, and $S_{\rm matter}$ is the action for matter fields. In general, the dynamics of gravity is defined by the measure weight $v$ and the curvature tensors (Riemann tensor, Ricci tensor and Ricci scalar) built with the metric and its derivatives ($\p_\mu$, ${}^\b\cD_\mu$, $\p_{q^\mu}$ or $\p_\mu^\g$, depending on the theory). The general structure of the Levi-Civita connection, the Ricci tensor, the Ricci scalar and the Einstein tensor in multifractional theories is
\ba
\Gamma^\rho_{\mu\nu}[\mathbf{D}] &:=& \frac12 g^{\rho\s}\left(\mathbf{D}_\mu g_{\nu\s}+\mathbf{D}_\nu g_{\mu\s}-\mathbf{D}_\s g_{\mu\nu}\right),\\
\cR_{\mu\nu}[\mathbf{D},\bar{\mathbf{D}}]&:=& \bar{\mathbf{D}}_\s \Gamma^\s_{\mu\nu}[\mathbf{D}]-\bar{\mathbf{D}}_\nu \Gamma^\s_{\mu\s}[\mathbf{D}]+\Gamma^\tau_{\mu\nu}[\mathbf{D}]\Gamma^\s_{\s\tau}[\mathbf{D}]-\Gamma^\tau_{\mu\s}[\mathbf{D}]\Gamma^\s_{\nu\tau}[\mathbf{D}]\,,\\
\cR[\mathbf{D},\bar{\mathbf{D}}]&:=& g^{\mu\nu}\cR_{\mu\nu}[\mathbf{D},\bar{\mathbf{D}}]\,,\\
\cG_{\mu\nu}[\mathbf{D},\bar{\mathbf{D}}]&:=& \cR_{\mu\nu}[\mathbf{D},\bar{\mathbf{D}}]-\frac12g_{\mu\nu}\cR[\mathbf{D},\bar{\mathbf{D}}]\,,
\ea
where $\mathbf{D}$ and $\bar{\mathbf{D}}$ denote generic derivative operators, not necessarily equal to each other due to symmetry requirements \cite{frc11} that go beyond the scope of this introductory review. When $\mathbf{D}=\bar{\mathbf{D}}$, we only write one argument in the curvature tensors, in particular, $\cR_{\mu\nu}[\mathbf{D}]$ and $\cR[\mathbf{D}]$. Furthermore, when $\mathbf{D}=\p$ we denote the standard Ricci tensor, Ricci scalar and Einstein tensors with the usual symbols
\be
R_{\mu\nu}=\cR_{\mu\nu}[\p]\,,\qquad R=\cR[\p]\,,\qquad G_{\mu\nu}=\cG_{\mu\nu}[\p]\,.
\ee

In the theories $T[\p^{\g(\ell)}]$ and $T[\B^{\g(\ell)}]$, the action has an extra integration over a length parameter $\ell$, possibly with a measure $\tau(\ell)$. For a single-scale geometry \cite{frc2,mf1},
\be
S=\frac{1}{2\k^2\ell_*}\int_0^{+\infty}\rmd\ell\,\tau(\ell)\int\rmd^Dx\,\sqrt{|g|}\,\cL_g^{(\ell)}+S_{\rm matter}\,.
\ee
The Lagrangians $\cL_g$ and $\cL_g^{(\ell)}$ and the findings in the literature are summarized in Tables \ref{tab4} and \ref{tab5}.

\begin{table}
\tbl{Characteristics of and topics in classical gravity in multifractional theories with integer-order operators. ``Diffeo'' stands for diffeomorphism. Empty cells correspond to topics not studied yet. Items with a tick \ding{51} indicate that a certain feature has been studied, while a question mark ``?'' indicates partial results.}
{\makebox[\textwidth][c]{
\begin{tabular}{|c||c|c|c|}\hline
  & $\bm{T_1}$ & $\bm{T_v}$ & $\bm{T_q}$ \\
{\bf Theory} & \makecell{{\bf ordinary derivatives}} & \makecell{{\bf weighted derivatives}} & $\bm{q}${\bf-derivatives}\\\hline\hline
Lagrangian $\cL_g$ & \makecell{$\vphantom{\displaystyle\int}R-\om(v)\p_\mu v\p^\mu v-2U(v)$\\ $\om$ and $U$ arbitrary \cite{fra2,frc11}} & \makecell{$\vphantom{\displaystyle\int}\cR[{}^\b\cD,\p]-\om(v)\cD_\mu v\cD^\mu v-2U(v)$\\ $\om$ and $U$ arbitrary \cite{frc11}} & \makecell{$\vphantom{\displaystyle\int}\cR[\p_q]-2\Lambda$\\ $\Lambda={\rm const}$ \cite{frc11}} \\\hline
Symmetries   & Broken diffeos \cite{frc11} & \makecell{$\bullet$ Without matter:\\ standard diffeos \cite{Calcagni:2016ivi}\\ $\bullet$ With matter:\\ broken diffeos \cite{frc11,Calcagni:2016ivi}} & \makecell{$q$-coordinates\\ diffeos \cite{frc11,Calcagni:2016ivi}} \\\hline
Big bang		 & \makecell{Bounce with exotic\\ matter \cite{fra2}} & \makecell{Bounce in vacuum \cite{frc11}} & \makecell{?\\ Bounce in vacuum\\ for $\a_{0,*}=0$ and\\ log oscillations \cite{frc11}}\\\hline
Inflation    & \makecell{?\\ Geometry can sustain\\ acceleration with\\ stiff matter \cite{fra2} $P=\rho$} & \makecell{Geometry alone\\ can sustain acceleration \cite{frc11}} & \makecell{Inflaton field\\ in mild slow-roll\\ required \cite{frc11,frc14}} \\\hline
Black holes  & & \makecell{Singularities not avoided \cite{Calcagni:2017ymp}} & \makecell{Singularities\\ not avoided \cite{Calcagni:2017ymp,Huang:2020lbv}} \\\hline
GWs: & \multicolumn{3}{c|}{} \\\hdashline
$\bullet$ dispersion relation & Standard \cite{qGW} & Standard \cite{qGW} & \ding{51} \cite{qGW} \\\hdashline
$\bullet$ luminosity distance & & & \ding{51} \cite{Calcagni:2019ngc} \\\hdashline
$\bullet$ stochastic background & & & \ding{51} \cite{Calcagni:2020tvw} \\\hline
Dark energy  & \makecell{?\\ $\bullet$ Geometry alone\\ cannot sustain\\ late-time acceleration \cite{fra2}\\
$\bullet$ Geometry can sustain\\ late-time acceleration\\ with mild slow-roll\\ matter \cite{Das:2018bxc} $P=w\rho$,\\ $-1/2<w<0$} & \makecell{Geometry alone can sustain\\ late-time acceleration\\ with $\a_{0,{\rm c}}\approx 4$ in FLRW \cite{frc11,Calcagni:2020ads}} & \makecell{Geometry alone\\ cannot sustain\\ late-time\\ acceleration \cite{Calcagni:2020ads}} \\\hline
\makecell{Alternative to\\ dark matter} & & & \\\hline
\end{tabular}\label{tab4}}}
\end{table}


\begin{table}
\tbl{Characteristics of and topics in classical gravity in multifractional theories with fractional operators. Empty cells correspond to topics not studied yet. Items with a question mark ``?'' indicate partial results.}
{\makebox[\textwidth][c]{
\begin{tabular}{|c||c|c|c|c|}\hline
  & $\bm{T[\p+\p^\g]}$ & $\bm{T[\p^{\g(\ell)}]}$ & $\bm{T[\B+\B^\g]}$ & $\bm{T[\B^{\g(\ell)}]}$ \\
{\bf Theory} & \multicolumn{2}{c|}{{\bf fractional derivatives}} & \multicolumn{2}{c|}{{\bf fractional d'Alembertian}}\\\hline\hline
\makecell{Lagrangian $\cL_g$\\ (one scale)} & \makecell{$\vphantom{\displaystyle\int}\cR[\cD]-2\Lambda$\\ $\Lambda={\rm const}$ \cite{mf1}} & \makecell{$\vphantom{\displaystyle\int}\cR[\p^{\g(\ell)}]-2\Lambda$\\ $\Lambda={\rm const}$ \cite{mf1}} & \makecell{$R-2\Lambda$\\ \!\!$\vphantom{\displaystyle\int}+\ell_*^2 G_{\mu\nu}(-\ell_*^2\B)^{\g-2}\,R^{\mu\nu}$\!\!\\ $\Lambda={\rm const}$ \cite{mf1}} & \makecell{$R-2\Lambda$\\\!\!$\vphantom{\displaystyle\int}+ G_{\mu\nu}\dfrac{(-\ell_*^2\B)^{\g(\ell)-1}-1}{\B}\,R^{\mu\nu}$\!\!\\ $\Lambda={\rm const}$ \cite{mf1}} \\\hline
Symmetries   & \makecell{Broken diffeos \cite{mf1}\\ $\bullet$ IR: ordinary\\ diffeos\\ $\bullet$ At plateaux\\ $\g\simeq{\rm const}$:\\ fractional diffeos} & \makecell{Fractional diffeos\\ of variable order} & \multicolumn{2}{c|}{Standard diffeos \cite{mf1}} \\\hline
\makecell{Black holes,\\ GWs, Inflation,\\ Dark matter} & & & & \\\hline
Dark energy & & & \makecell{?\\ Reproduces IR nonlocal\\ gravity for $\g\to 0,1$ \cite{mf1}} & \\\hline
\end{tabular}\label{tab5}}}
\end{table}


\subsection{About cosmology}

Studies on late-time acceleration have been carried out on a homogeneous and isotropic background, in particular, a flat Friedmann--Lema\^itre--Robertson--Walker (FLRW) metric. 

In the theory $T_1$ with ordinary derivatives, the problem has been considered mainly with exotic dark-energy components, in flat \cite{fra2,Karami:2012ra,Lemets:2012fp,Chattopadhyay:2013mwa,Jawad:2016piq,Sadri:2018lzz,Ghaffari:2019qcv,Mamon:2020ocb} as well as non-flat FLRW \cite{Maity:2016dfv,Debnath:2019isy}. In these cases, late-time acceleration is possible, although the theoretical motivation is no more robust than in general relativity. It is indeed possible to realize dark energy with an ordinary fluid with mildly negative barotropic index $w$ \cite{Das:2018bxc}, but it is not clear whether a scalar field with such properties would need fine tuning just like quintessence; hence the question mark in the table. 

The conclusion reached in the theory $T_q$ is that a cosmological constant or exotic fluids are required at late times to sustain acceleration \cite{Calcagni:2020ads}. Thus, while in $T_1$ such fluids are optional and there is still the possibility to get acceleration with conventional matter, in $T_q$ this option seems barred.

In contrast with these theories, in the theory $T_v$ geometry can sustain acceleration without the need of matter, provided an ultra-IR regime exists \cite{frc11,Calcagni:2020ads}. This intriguing scenario has been tested with late-time data but it awaits a more complete analysis.

The dark-matter row in Tables \ref{tab4} and \ref{tab5} refers to the possibility of explaining galaxy rotation curves within the multifractional paradigm, without invoking a dark matter component. To date, this possibility has not been explored.


\subsection{About theories with fractional operators (II)}

Inspired by Ho\v{r}ava--Lifshitz gravity \cite{Calcagni:2009qw}, fractional derivatives and integrals have been invoked since the earliest papers on the multifractional paradigm \cite{fra1,frc1,frc2}, but it was only very recently that multifractional theories with fractional derivatives have been constructed explicitly \cite{mf1}. This is the reason why Table \ref{tab5} is emptier than the others: there has been little time to develop the phenomenology of these theories. Also, originally only one theory with fractional operators was envisaged (it was called $T_\g$ in Ref.\ \citen{revmu}), while now we can recognize at least four.


\section{QFT and quantum gravity in multifractional theories}\label{sec4}

In this section, we summarize the status of multifractional theories as quantum field theories of matter and gravity (Tables \ref{tab6} and \ref{tab7}).
\begin{table}
\tbl{QFT of matter and gravity in multifractional theories with integer-order operators. Empty cells correspond to topics not studied yet. Items with a tick \ding{51} indicate that a certain feature has been studied, while a question mark ``?'' indicates partial results.}
{\makebox[\textwidth][c]{
\begin{tabular}{|c||c|c|c|}\hline
  & $\bm{T_1}$ & $\bm{T_v}$ & $\bm{T_q}$ \\
{\bf Theory} & \makecell{{\bf ordinary}\\ {\bf derivatives}} & \makecell{{\bf weighted}\\ {\bf derivatives}} & $\bm{q}${\bf-derivatives}\\\hline\hline
Scalar field & Refs.\ \citen{fra1,frc2} & Refs.\ \citen{frc6,frc13} & Refs.\ \citen{frc12,frc13} \\\hline
\makecell{Gauge fields and\\ Standard Model} & Refs.\ \citen{MSBR,revmu} & Refs.\ \citen{frc8,frc13} & Ref.\ \citen{frc13} \\\hline
Quantum gravity & \makecell{Discussed\\ using scalar\\ QFT \cite{fra1,frc2}} & \makecell{Discussed\\ using scalar\\ QFT \cite{frc9}} & Discussed using scalar QFT \cite{frc9} \\\hline
Unitarity & No \cite{fra2,revmu} & Yes & Yes \\\hline
\makecell{Improved\\ power-counting\\ renormalizability} & \makecell{Yes \cite{fra1,fra2,frc2,revmu}\\if $\vphantom{\displaystyle\int}\a\leq\frac{2}{D}$\\ In $D=4$: $\a\leq\frac12$} & No \cite{frc9} & Inconclusive \cite{frc2,revmu} \\\hline
\makecell{Improved\\ perturbative\\ renormalizability} & & No \cite{frc9,revmu} & \makecell{No in deterministic view \cite{frc9,revmu}\\ ? in stochastic view \cite{revmu}}\\\hline
\end{tabular}\label{tab6}}}
\end{table}

\begin{table}
\tbl{QFT of matter and gravity in multifractional theories with fractional operators. Empty cells correspond to topics not studied yet. Items with a tick \ding{51} indicate that a certain feature has been studied, while a question mark ``?'' indicates partial results.}
{\makebox[\textwidth][c]{
\begin{tabular}{|c||c|c|c|c|}\hline
  & $\bm{T[\p+\p^\g]}$ & $\bm{T[\p^{\g(\ell)}]}$ & $\bm{T[\B+\B^\g]}$ & $\bm{T[\B^{\g(\ell)}]}$ \\
{\bf Theory} & \multicolumn{2}{c|}{{\bf fractional derivatives}} & \multicolumn{2}{c|}{{\bf fractional d'Alembertian}}\\\hline\hline
Scalar field & \multicolumn{4}{c|}{Ref.\ \citen{mf1}} \\\hline
\makecell{Gauge fields and\\ Standard Model} & & & & \\\hline
Quantum gravity & \multicolumn{4}{c|}{Ref.\ \citen{mf1}} \\\hline
Unitarity & & & \multicolumn{2}{c|}{\makecell{$\bullet$ Yes \cite{mf1} if $n\in\mathbb{N}$, $-2n<\g<1-2n$.\\ $\ds>0$ only if $n=0$: $\vphantom{\displaystyle\int}0<\g<1$\\ $\bullet$ No \cite{mf1} when $\vphantom{\displaystyle\int}\g>1$}} \\\hline
\makecell{Improved\\ power-counting\\ renormalizability} & \multicolumn{4}{c|}{Yes \cite{frc2,mf1} if $\g\geq\dfrac{D}{2}$.\quad In $D=4$: $\g\geq 2$} \\\hline
\makecell{Improved\\ perturbative\\ renormalizability} & & & \multicolumn{2}{c|}{\makecell{$\bullet$ Yes at 1-loop \cite{mf1} if $n\in\mathbb{N}$,\\ $\vphantom{\displaystyle\int}\dfrac{D}{2}-n\neq\g\neq\dfrac{D}{4}-\dfrac{n}{2}$\\ $\bullet$ In $D=4$ when $\g>1$: $\vphantom{\displaystyle\int}\g\neq2$}} \\\hline
\end{tabular}\label{tab7}}}
\end{table}


\subsection{About quantum gravity}

The quantum-gravity row in Tables \ref{tab6} and \ref{tab7} refers to the discussion of the theory as an independent perturbative QFT of gravity. Other papers dealt with the relationship and similarities between multifractional theories and other theories of quantum gravity \cite{frc2,revmu,Amelino-Camelia:2017pdr,Calcagni:2017jtf,Calcagni:2016ivi,Arzano:2011yt,Calcagni:2012cn}.


\subsection{About renormalizability}

The first three papers on multiscale field theories \cite{fra1,fra2,fra3} considered a non-factorizable Lebesgue--Stieltjes measure. In these spacetimes, the momentum transform \cite{fra2,fra3} is different with respect to the transform in terms of Bessel functions of the multifractional theories $T_1$ and $T_v$ \cite{frc3}. Many of the results for the scalar field in \cite{fra1,fra2,fra3} look similar to those for the scalar field in $T_1$ but, in fact, there are some differences \cite{frc2}. However, the power-counting argument is the same and also the equations of motion.

The theory $T_1$ is difficult to handle due to the absence of action symmetries, a non-self-adjoint kinetic operator, and an issue with unitarity \cite{revmu}. For these reasons, its development as a QFT has not gone beyond some basic results at the tree level \cite{frc2}.

In the theory $T_v$ with weighted derivatives, perturbative QFT is unviable in the presence of nonlinear interactions, as in the case of an isolated scalar field \cite{frc9,frc13}. However, this problem does not arise for the full Standard Model due to the presence of an integer frame where the theory is made formally equivalent to the ordinary Standard Model in all sectors \cite{frc13}.

Power-counting renormalizability is determined by the superficial degree of divergence, which is \cite{frc2,mf1}
\be\label{dmax}
\de \leq (D\a-2\g)L\,,
\ee
where $L$ is the number of loops in a one-particle-irreducible Feynman diagram. Replacing $\g=1$ for the theory $T_1$, $\g=\a$ for the theory $T_q$ (and $T_\a$, not reported in the tables), and $\a=1$ for the theories with fractional operators and ordinary measure, one gets the results reported in Tables \ref{tab6} and \ref{tab7}. 

In particular, while in the deterministic view the superficial degree of divergence of $T_q$ and $T_\a$ is the same as a standard QFT, in the stochastic view it is possible that the stochastic fluctuations of the measure render spacetime fuzzy at the scale $\ell_*$ and the concept of coincident spacetime points loses meaning \cite{revmu}. Whether this leads to an improved perturbative renormalizability is not clear, and the power-counting argument is not conclusive.

The case of the theory $T_v$ is also delicate because the momentum-space basis $\bE(k,x)$ carries a measure weight that changes the power counting and, eventually, renormalizability is not improved because momentum integrals have the same degree of divergence than a standard QFT \cite{frc9,revmu}.

Regarding the theories with fractional operators, unitarity and renormalizability with fractional derivatives have not been studied yet, apart from power-counting renormalizability. More is known for the cases with fractional d'Alembertian. The theory $T[\B+\B^\g]$ cannot be at the same time unitary and perturbative renormalizable, since the range of values of $\g$ for which unitarity is respected never intersects the ranges for which the theory has improved renormalizability.


\subsection{About unitarity}

Based on the nonconservation of Noether currents, in previous papers it was claimed that the multifractional theories $T_1$, $T_v$ and $T_q$ are not unitary \cite{fra2,frc6,revmu}. This was not felt as a problem at least for $T_v$ and $T_q$ because one could reformulate these theories in the integer frame as unitary models and, somehow, control the loss of unitarity in the fractional frame. However, here we show that at least $T_v$ and $T_q$ are indeed unitary. To do so, we work in the fractional (physical) frame and check the property of reflection positivity in the Euclidean version of the theory. The details of the procedure can be found, e.g., in Refs.\ \citen{mf1,Tri17} and amounts to show that the scalar product of field functionals $\vp$ defined through the Green's function is positive definite.

In Euclidean position space with coordinates $x_1,x_2,\dots, x_D$, one defines a reflection operation ${\rm R}$ such that spatial coordinates are unchanged while ${\rm R} x_D=-x_D$. For any test function $\vp$ chosen in an appropriate functional space, for a generic multifractional theory we have to show that 
\be
\left({\rm R}\vp,\vp\right):=\int\rmd^Dq(x)\,\rmd^Dq(x')\,\vp^*(x)\,G({\rm R}x,x')\,\vp(x')\geq 0\,,
\ee
where the Green's function $G$ is given by Eq.\ \Eq{scw}. We choose a charge-distribution-type of test functions, which on a multifractional spacetime is $\vp(x)=\sum_{i=1}^N b_i\de^D[q(x)-q(x^{(i)})]$, where $b_i\in\mathbb{C}$. Calling $\a_{\bm{k}}[\bm{x}^{(i)}]:=b_i\prod_{\mu=1}^{D-1}\bE[k_\mu,x_\mu^{(i)}]$, we have
\ba
\left({\rm R}\vp,\vp\right) &=& \sum_{i,j}b_i^*b_j G[{\rm R}x^{(i)},{x'}^{(j)}]\nonumber\\
					 &=&-\sum_{i,j}\int_{-\infty}^{+\infty}\rmd^{D-1}\bm{p}(\bm{k})\a^*_{\bm{k}}[\bm{x}^{(i)}]\a_{\bm{k}}[\bm{x'}^{(j)}]\nonumber\\
					&&\qquad\times\int_{-\infty}^{+\infty}\rmd p_D(k_D)\,\frac{\bE_D[k_D,x_D^{(i)}]\bE_D[k_D,{x_D'}^{(i)}]}{\cK(k)}\nonumber\\
					 &=:& \sum_{i,j}\int_{-\infty}^{+\infty}\rmd^{D-1}\bm{p}(\bm{k})\,\a^*_{\bm{k}}[\bm{x}^{(i)}]I^{ij}\a_{\bm{k}}[\bm{x'}^{(j)}]\,,
\ea
where we used Eq.\ \Eq{facte}. Reflection positivity holds if $I^{ij}\geq 0$. 

In the theory $T_1$, the basis $\bE(k,x)$ is made of Bessel functions and the calculation of $I^{ij}$ becomes involved. We will not consider this case here but we note that the no-unitarity arguments of Refs.\ \citen{fra2,revmu} remain valid, since there is no integer frame here.

The basis $\bE$ in the theory $T_v$ is given by Eq.\ \Eq{tve}. Adding a mass term to the kinetic operator $\cK=\B+(\p_\mu v/v)\p^\mu-m^2$, one has $\cK\bE(k,x)=-(k^2+m^2)\bE(k,x)$. In Euclidean space, $k^2+m^2=\sum_\mu k_\mu^2+m^2=:k_D^2+\om_k^2$, where $\om_k^2:=|\bm{k}|^2+m^2$ so that, denoting $r_{ij}:=x_D^{(i)}+{x_D'}^{(j)}$,
\ba
I^{ij}&=&-\int_{-\infty}^{+\infty}\rmd k_D\,w_D(k_D)\,\frac{\bE_D[k_D,x_D^{(i)}]\bE_D[k_D,{x_D'}^{(j)}]}{\cK(k)}\nonumber\\
&=& \frac{1}{\sqrt{v_D[x_D^{(i)}]v_D[{x_D'}^{(i)}]}}\int_{-\infty}^{+\infty}\frac{\rmd k_D}{2\pi}\,\frac{\rme^{\rmi k_Dr_{ij}}}{k_D^2+\om_k^2}\nonumber\\
&=& \frac{\rme^{-\om_k |r_{ij}|}}{2\om_k\sqrt{v_D[x_D^{(i)}]v_D[{x_D'}^{(i)}]}}>0\,.
\ea
Therefore, the theory obeys reflection positivity and, by analytic continuation to Lorentzian signature, it is unitary. This is in agreement with the fact that the S-matrix in the quantum mechanics of $T_v$ is unitary \cite{Calcagni:2012kf}.

The basis $\bE$ in the theory $T_q$ is given by Eq.\ \Eq{tqe} and $\cK\bE(k,x)=-[p_\mu (k^\mu) p^\mu(k^\mu)+m^2]\bE(k,x)$ with a mass term. In Euclidean space, $p^2+m^2=\sum_\mu p_\mu^2+m^2=p_D^2+\om_p^2$, where $\om_p^2:=|\bm{p}|^2+m^2$, implying
\be
I^{ij}= \int_{-\infty}^{+\infty}\frac{\rmd p_D}{2\pi}\,\frac{\rme^{\rmi p_Dq_{ij}}}{p_D^2+\om_p^2}= \frac{\rme^{-\om_p |q_{ij}|}}{2\om_p}> 0\,,
\ee
where $q_{ij}:=q_D[x_D^{(i)}]+q_D[{x_D'}^{(j)}]$. Thus, also $T_q$ is unitary.


\section{What next?}\label{sec5}

We conclude by listing some of the topics to be explored in the near future.
\begin{itemize}
\item With the study of multifractional theories with integer-order derivatives almost complete, attention has been recently shifted to the theories with fractional operators \cite{mf0,mf1}. Unitarity and one-loop renormalizability of the theory $T[\p+\p^\g]$ is an open question, even if we expect similar problems than for the theory $T[\B+\B^\g]$. As a start, one could employ the methods of Ref.\ \citen{mf1} to check these properties for the no-scale theory $T[\p^\g]$.
\item The theory $T[\B^{\g(\ell)}]$ with variable-order fractional operators could avoid the renormalizability-versus-unitarity problem of $T[\B+\B^\g]$, but the details of how to manipulate the integration over $\ell$ in calculations have not been worked out.
\item As noted in Ref.\ \citen{revmu}, the theories $T_\a[\p+\p^\a]$ and $T_\a[\B+\B^\a]$ with multifractional measure could be akin to $T_q$. Studying the correspondence \Eq{cong} might help to understand how to develop these theories to the point of extracting observational predictions.
\item The big-bang problem has been cursorily touched upon in Ref.\ \citen{frc11} for the theories with integer-order operators and a bounce may be possible in $T_v$ and $T_q$ without invoking exotic matter. It would be interesting to develop more detailed bouncing models.
\item To date, black holes and cosmology in theories with fractional operators are still virgin territory.
\item There are promising signs that the theories $T_1$ and $T_v$ can sustain inflation with or without matter fields \cite{frc11}. However, no study of primordial scalar, vector and tensor perturbations and of the corresponding spectra has been carried out.
\item The problem of dark energy has been explored extensively for the theory $T_1$, but only one paper pointed out a scenario with a conservatively realistic fluid component \cite{Das:2018bxc}. This has been done with a power-law measure weight $v=a^m$, where $a(t)$ is the scale factor, and without trying to realize the same equation of state with a scalar field. Therefore, it remains to be seen how a multifractional weight $v= 1+a^m+\dots$ would modify these results, or whether a scalar field would be subject to a fine tuning on the initial conditions similarly to quintessence in general relativity. Moving to scenarios with fractional operators, the theory $T[\B+\B^\g]$ with fractional d'Alembertian could have an important application in explaining the late-time acceleration of the universe \cite{mf1}. In fact, in the limits $\g\to 0,1$ it can reproduce, unify and theoretically justify classical models of IR modifications of gravity with Lagrangian
\be
\cL= R+ c_0R \B^{-n_0} R+c_2 R_{\mu\nu} \B^{-n_2} R^{\mu\nu}\,,
\ee
where $c_{0,2}$ are constants and $n_{0,2}$ classify different scenarios: the $n_0=1=n_2$ model \cite{Barvinsky:2003kg,Barvinsky:2005db,Barvinsky:2011hd,Barvinsky:2011rk}, the $n_0=0$, $n_2=1$ model \cite{Ferreira:2013tqn,Nersisyan:2016jta}, the $n_0=0$, $n_2=2$ model \cite{Cusin:2015rex,Zhang:2016ykx}, and the $n_0=2$, $n_2=0$ model \cite{Maggiore:2014sia,Belgacem:2020pdz}.
\item The problem of finding alternatives to dark matter has not been considered in any multifractional theory, with integer-order or fractional operators.
\end{itemize}
In our opinion, the value of the multifractional paradigm can be appreciated especially when phenomenological explorations, for instance in cosmology, are pursued with the goal of offering scenarios with less fine tuning and less exotic matter components than in general relativity. We hope that this short review will stimulate the reader in that direction.




\end{document}